\documentstyle[amssymb,draft,11pt,epsf,palatino]{nature-pvd}

\def\simlt{\mathrel{\hbox{\rlap{\hbox{\lower4pt\hbox{$\sim$}}}\hbox{$<$}}}}
\def\simgt{\mathrel{\hbox{\rlap{\hbox{\lower4pt\hbox{$\sim$}}}\hbox{$>$}}}}

\def\ale{\mathrel{\hbox{\rlap{\hbox{\lower4pt\hbox{$\sim$}}}\hbox{$<$}}}}
\def\age{\mathrel{\hbox{\rlap{\hbox{\lower4pt\hbox{$\sim$}}}\hbox{$>$}}}}

\def\kms{km\,s$^{-1}$}
\def\msun{M$_{\odot}$}
\def\g1256{1255--0}

\def\spose#1{\hbox to 0pt{#1\hss}}
\newcommand\lsim{\mathrel{\spose{\lower 3pt\hbox{$\mathchar''218$}}
     \raise 2.0pt\hbox{$\mathchar''13C$}}}
\newcommand\gsim{\mathrel{\spose{\lower 3pt\hbox{$\mathchar''218$}}
     \raise 2.0pt\hbox{$\mathchar''13E$}}}

\begin{document}
\gdef\ha{H$\alpha$}
\gdef\ew{${\rm EW}({\rm H}\alpha)$}
\gdef\msun{M$_{\odot}$}
\gdef\kms{km\,s$^{-1}$}

\twocolumn

\title{\Large \bf {A massive galaxy in its core formation phase three billion years after the Big Bang}}

%Uncovering a Compact Mode of Star Formation in the Early Universe
% FOR SUBMISSION - REMOVED \small FROM NEXT:
\author{{Erica Nelson}\affiliation[1]
  {{Astronomy Department, Yale University,  New Haven, CT, USA}
},
   {{Pieter van Dokkum}}$^1$,   
   {Marijn Franx}\affiliation[2]
   {{Leiden Observatory, Leiden University, Leiden, The Netherlands}},
{Gabriel Brammer}\affiliation[3]
 {{Space Telescope Science Institute, 3700 San Martin Drive,
Baltimore, MD 21218, USA}},
{Ivelina Momcheva}$^1$,
{Natascha M.\ F\"orster Schreiber}\affiliation[4]
{{Max-Planck Institut f\"ur Extraterrestrische Physik,
Giessenbackstrasse, D-85748 Garching, Germany}},
{Elisabete da Cunha}\affiliation[5]
{{Max Planck Institute f\"ur Astronomie, K\"onigstuhl 17,
D-69117, Heidelberg, Germany}},
 {Linda Tacconi}$^4$,
{Rachel Bezanson}\affiliation[6]
{{Steward Observatory, University of Arizona, 933 N.\
Cherry Avenue, Tucson}},
{Allison Kirkpatrick}\affiliation[7]
{{Department of Astronomy, University of Massachusetts, Amherst, MA 01002,
USA}},
{Joel Leja}$^1$,
{Hans-Walter Rix}$^5$,
{Rosalind Skelton}\affiliation[8]
{{South African Astronomical Observatory, P.O.\ Box 9,
Observatory, 7935, South Africa}},
{Arjen van der Wel}$^5$,
{Katherine Whitaker}\affiliation[9]
{{Astrophysics Science Division, Goddard Space Center,
Greenbelt, MD 20771, USA}},
{Stijn Wuyts}$^4$
   {}\vspace{0.4cm}
}
%\date{\today}{}
\headertitle{Compact Star Formation}
\mainauthor{Nelson et al.}

\summary{Most massive galaxies are thought to have 
formed their dense stellar cores at early cosmic epochs.
\cite{daddi:05,oser:10,vandokkum:14}
However, cores in their formation phase have not yet been observed. 
Previous studies have found galaxies with high gas velocity 
dispersions \cite{tacconi:08} or small apparent sizes 
\cite{toft:14,barro:14a,williams:14} but so far no objects have been 
identified with both the stellar structure and the gas dynamics of a forming core.
Here we present a candidate core
in formation 11 billion years ago, at $z=2.3$.
GOODS-N-774 has a stellar mass of $1.0\times 10^{11}$\,M$_{\odot}$,
a half-light radius of 1.0\,kpc,
and a star formation rate of $90^{+45}_{-20}$\msun/yr.
The star forming gas has a velocity dispersion $317\pm30$\,km/s, 
amongst the highest ever measured. 
It is similar to the stellar velocity dispersions of the putative 
descendants of GOODS-N-774, compact quiescent galaxies at 
$z\sim 2$\cite{bezanson:13,vandokkum:09b,vandesande:13,belli:14}
and giant elliptical galaxies in the nearby Universe.
Galaxies such as GOODS-N-774 appear to be rare; however,
from the star formation rate and size of the galaxy
we infer that many star forming cores may be heavily
obscured, and could be missed
in optical and near-infrared surveys.}

\maketitle 

%%%%%%%%%%%%%%%%%%%%%%%%%%%%%%%%%%%%%%%%%%%%%%%%%%%%%%%%%%%%%%%%%%%
%\onecolumn uncomment for submisssion

%% uncomment for submisssion ---
%\noindent
%{\bf
% move summary here for submission
%}
%\clearpage 
% -------

\vspace{-0.0cm}
We identified the candidate forming core, GOODS-N-774, using the
3D-HST catalogs in the five CANDELS fields.\cite{skelton:14}
GOODS-N-774 has a circularized effective radius 
$r_e=1.0$\,kpc from HST F160W WFC3 imaging;\cite{vanderwel:14} a
stellar mass of 
$1.0 \times 10^{11}M_\odot$ \cite{skelton:14,kriek:09}; 
rest-frame $UVJ$ 
colors consistent with a star-forming galaxy; %\cite{wuyts:11b}
and a MIPS 24\,$\mu$m flux of $104\,\mu$Jy.
Fig.\ 1 shows the stellar mass density profile derived from 
the observed $H_{160}$ surface brightness
profile corrected for the HST PSF.\cite{szomoru:10}
The surface density profile is strikingly similar to the average profile of
massive quiescent galaxies at $z\approx 2$ (red line),
and much more concentrated
than the average profile
of massive star forming galaxies at that redshift
(light blue).\cite{vanderwel:14} % Wuyts 11, Genzel 06
\noindent
\begin{figure}[h]
\vspace{9.65cm}
\end{figure}
%
%\vspace{3cm}\\
\noindent

The near infrared spectrum of GOODS-N-774 is shown in Fig.\,2. 
The continuum is clearly
detected, along with emission lines that we identify as
H$\alpha$ and [N\,{\sc ii}] redshifted to $z=2.300$. 
The gas velocity dispersion is
$\sigma=317\pm30$\,km/s, equivalent to
a FWHM\,$\approx 750$\,km/s.
Typically, objects with such large linewidths are mergers or
dominated by active galactic nuclei
(AGN).\cite{tacconi:08} If the line emission in GOODS-N-774 is partially or 
largely due to the presence of an AGN, its velocity dispersion, size, 
and stellar mass measurements would not be reliable.

There is no evidence for the presence of an active nucleus in 
GOODS-N-774. It is not detected in the deep Chandra 2\,Ms X-ray data 
in GOODS-North with an upper limit of 
$L_X<1.2\times 10^{42}$\,ergs\,s$^{-1}$. 
While an AGN cannot be conclusively ruled out, this upper 
limit is consistent with the star formation rate of the galaxy.
Also, the galaxy has line ratios [O\,{\sc iii}]/[O\,{\sc ii}]$=0.7\pm0.5$,
[O\,{\sc iii}]/H$\beta=1.2\pm0.9$,
and [NII]/\ha\,$=0.4\pm0.1$, indicating a low ionization state of the
gas. Therefore stellar photoionization, and hence ultimately star
formation, is the likely origin of the line emission.
Finally, the observed
infrared SED, shown in Fig.\ 3, requires strong PAH emission to 
simultaneously explain the MIPS 24\,$\mu$m and Herschel
data, effectively ruling out the presence of a dominant AGN.
We quantified this by fitting composite SEDs with varying
AGN contributions.\cite{kirkpatrick:12}
The best fit is obtained for
a pure star-forming template with 0\,\% AGN contribution
(see Fig.\ 3).

% forster-schreiber:11a
We infer that 
GOODS-N-774's line width is among the highest measured for a
normal star forming galaxy at high redshift, as shown in Extended Data Fig. 1.
If the gas is in a disk, it is rotating with 
a velocity of $v_c\approx550$\,km/s, or $v_c\approx680$\,km/s 
after correcting for inclination.
The observed gas velocity dispersion of 317\,km/s is similar to the median
stellar velocity dispersion of 304\,km/s in 
a sample of quiescent galaxies at
$z=1.5-2.2$ with median mass $1.9\times10^{11}$\msun\
\cite{bezanson:13,vandokkum:09b,vandesande:13,belli:14} (see Fig.\ 4).
\vspace{1cm}

\begin{figure}[htbp]
\epsfxsize=7.8cm
\epsffile{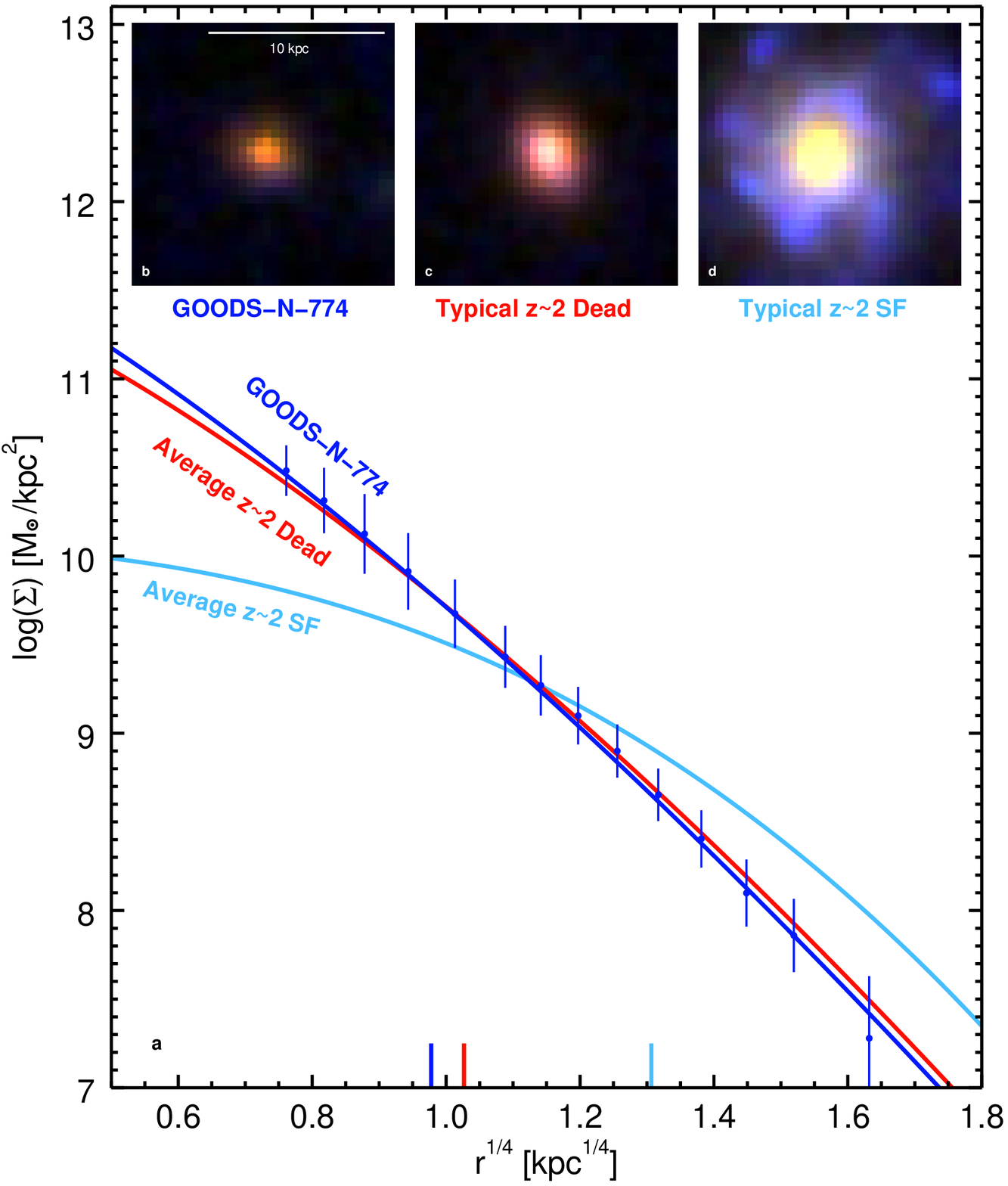}
%\noindent 
{\small \bf {\textsf{Figure 1}} $|$ } 
{\small \bf {\textsf{Structural properties of GOODS-N-774.}}}
{\small
\textbf{a}, Surface density profile of GOODS-N-774 (blue line),
as derived from deep WFC3 $H_{160}$ imaging.  Error bars are s.d.
The galaxy has a mass of $1.0\times 10^{11}\,M_{\odot}$ and an effective
radius $r_e=1.0$\,kpc. The light blue curve shows the average profile of 67
star forming galaxies at $1.9<z<2.1$ with
$10.9<\log(M_{\rm stellar})<11.2$.\cite{vanderwel:14,skelton:14}
The red curve shows the average profile of 24 quiescent galaxies with
the same mass and redshift selection criteria.  \textbf{b-d}, Color images show
GOODS-N-774, a typical quiescent galaxy, and a typical star forming
galaxy. Vertical bars indicate
effective radii.
GOODS-N-774's structure is similar to that of
massive quiescent galaxies.
% 97 words
}
\end{figure}

\begin{figure}[h]
%\vspace{0.2cm}
\epsfxsize=7.8cm
\epsffile{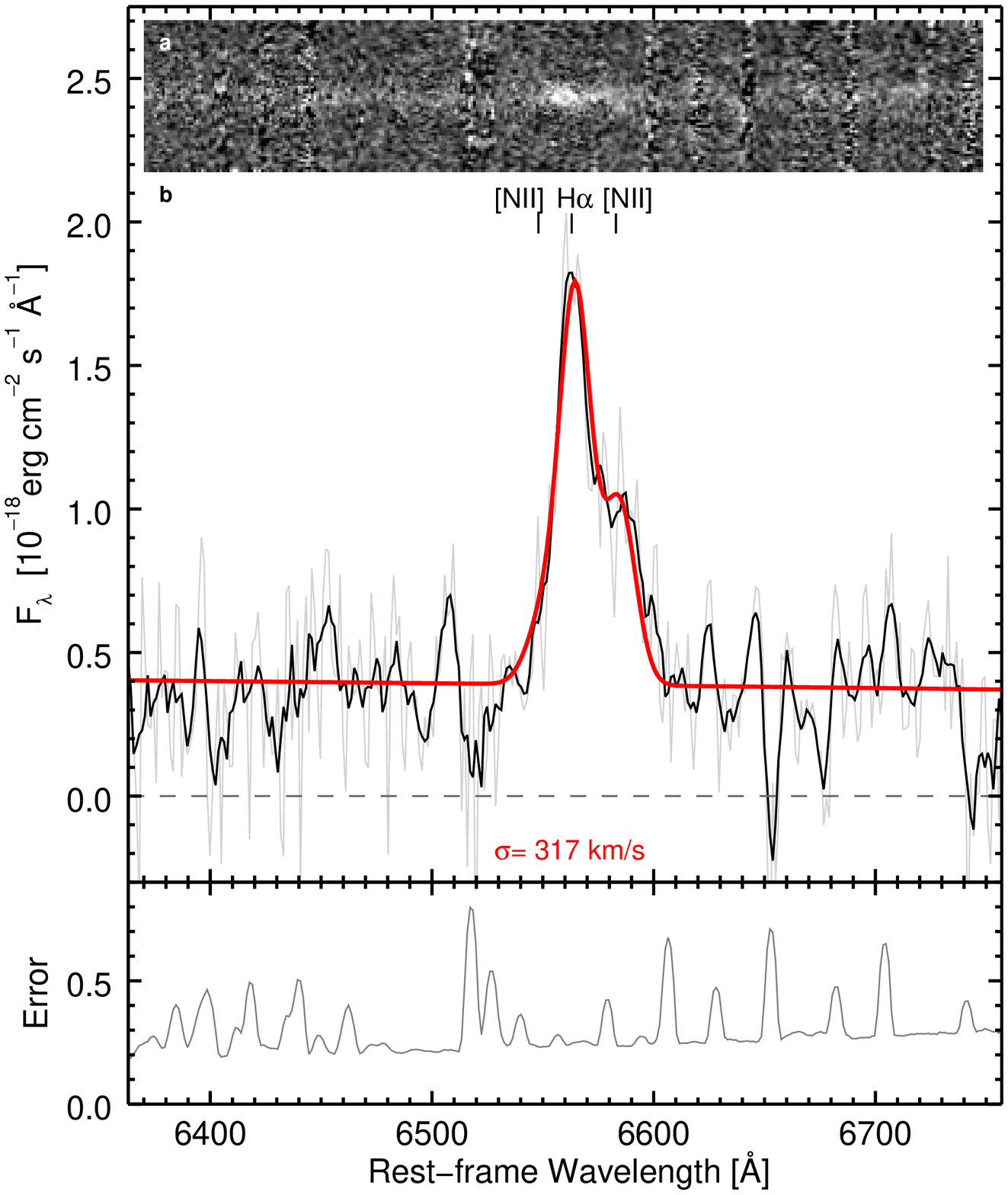}
{\small \bf {\textsf{Figure 2}} $|$ }
{\small \bf {\textsf{Velocity dispersion of GOODS-N-774.}}}
{\small
NIR spectrum in 2D (\textbf{a}) and 1D (\textbf{b}) obtained with NIRSPEC on Keck.
The grey curve is at the original resolution;
the black curve shows the spectrum smoothed with a 
20\,\AA\, boxcar. The best-fit gaussians to the  \ha\, $\lambda6563$\AA\, 
and [N\,{\sc ii}] $\lambda\lambda6548,6584$\AA\,
emission lines are shown in red. 
The velocity dispersion is $317\pm 30$\,km/s,
equivalent to an inclination-corrected circular velocity 
of $v_c\approx680$\,km/s if the gas is rotating in a disk.
The rest-frame equivalent width of H$\alpha$ is $66 \pm 8$\,\AA\, and 
its luminosity is $(3.4\pm{}0.4) \times 10^{42}$\,ergs/s.
}
\end{figure}

The inferred dynamical mass
is $1.5\times10^{11}$\msun, roughly twice the stellar mass, 
suggesting
a gas fraction of $\lesssim 50$\,\%.
In Fig.\ 4  we explicitly compare the dynamical and
structural properties of GOODS-N-774 to galaxies in
the Sloan Digital Sky Survey (SDSS) and to quiescent 
galaxies at $z\sim 2$. The galaxy
has a much smaller size and a higher velocity dispersion than
SDSS galaxies of the same total dynamical mass. Its properties
are very similar to those of
the samples of quiescent galaxies at $z\sim 2$
that have been assembled over the past few years, and we infer
that we have identified a star forming example of
galaxies in this region of parameter space.

The H$\alpha$ luminosity is $(3.4 \pm 0.4) \times 10^{42}$\,ergs/s,
which implies a minimum star formation rate
(with no reddening correction) of $\sim
16$\,M$_{\odot}$/yr  for a Chabrier
initial mass function.\cite{kennicutt:98araa,chabrier:03}
The red color of the galaxy ($R_{606} - H_{160} = 4.2$)
and the fact that it is detected with MIPS and Herschel
suggest that the actual, dust-corrected
star formation rate is much higher. The 24\,$\mu$m
flux alone indicates a star formation rate of
135\,M$_{\odot}$/yr.\cite{wuyts:11} % wuyts:08
Fitting the 24\,$\mu$m -- 500\,$\mu$m data
with empirical composite star forming SEDs\cite{kirkpatrick:12}
or theoretical models\cite{dacunha:08} gives slightly lower
values than the 24\,$\mu$m data alone, and we infer that
the  star formation rate is $90^{+45}_{-20}$\,\msun/yr.
This confirms that the
star formation is highly obscured, with $\sim 3$
magnitudes of extinction
toward H$\alpha$ and $L(IR)/L(UV) \gtrsim 200$.
\vspace{1cm}

GOODS-N-774 has a specific star formation 
rate of  $\sim 1\times10^{-9}$/yr, which places it on the
star forming sequence at $z=2.3$.\cite{wuyts:11}
If the galaxy had a constant star formation rate prior to
the epoch of observation its mass
was built up over a period of
$\sim 1$\,Gyr since $z\sim 3.3$. Although short compared
to the age of the Universe at $z=2.3$, this build-up phase
is $\sim 200$ dynamical times,  longer than expected from
the Kennicutt-Schmidt law.\cite{kennicutt:98}
This suggests that the
galaxy had a higher star formation rate in the past, or that the
star formation rate has been throttled by the gas accretion rate
onto the halo:
a galaxy with a stellar mass of $M_*=1.0\times10^{11}$\msun\, 
would have a baryonic accretion rate of $\sim 60-120$\msun/yr,
\cite{dekel:13} in good agreement with with the observed star formation rate.

The gas in a galaxy such as this, growing via rapid star formation in
a deep potential, should be rapidly enriched with metals and we would
thus expect it to exhibit a high gas-phase metallicity. This is
consistent with what we observe: the galaxy has [NII]/\ha\,$=0.4\pm0.1$,
which implies a high metallicity (12+log$(\frac{O}{H})\sim9.05$,
although the conversion\cite{maiolino:08} is somewhat uncertain).
After the star formation phase the gas is probably
heated and/or expelled.\cite{oser:10,dekel:13} 
The quiescent core that remains will 
then likely evolve into a giant elliptical galaxy\cite{oser:10,vandokkum:14}
with a central stellar metallicity that is similar
to the gas-phase metallicity 
of the star-forming core at high redshift.\cite{leja:13}

Galaxies such as GOODS-N-774 are rare. Candidate
compact star forming
galaxies with less extreme properties have been identified
in fairly large numbers,\cite{toft:14,barro:14a} but in the 
$\sim900$\,arcmin$^2$ of the five
3D-HST/CANDELS fields
there are only three objects at $2<z<3$ with 24\,$\mu$m flux
$\geq 100\,\mu$Jy, a high central mass density
($\log(M/M_{\odot})(r<1$\,kpc)\,$\geq 10.5$; see ref.\ 7), 
and a concentrated stellar distribution ($r_e\leq 1$\,kpc).
We observed all three galaxies with Keck and GOODS-N-774 is
the only confirmed candidate: GOODS-S-5981\cite{barro:14a} has
a narrow line width, whereas COSMOS-8388 is 
difficult to interpret because it has an active nucleus. 
The number density
we infer is $\sim 10^{-6}$\,Mpc$^{-3}$ 
(including all three candidates), compared to
 $\sim10^{-4}$\,Mpc$^{-3}$
for the overall population of galaxies with dense cores
at $z\sim2$.\cite{vandokkum:14}

\begin{figure}
\epsfxsize=7.8cm
\epsffile{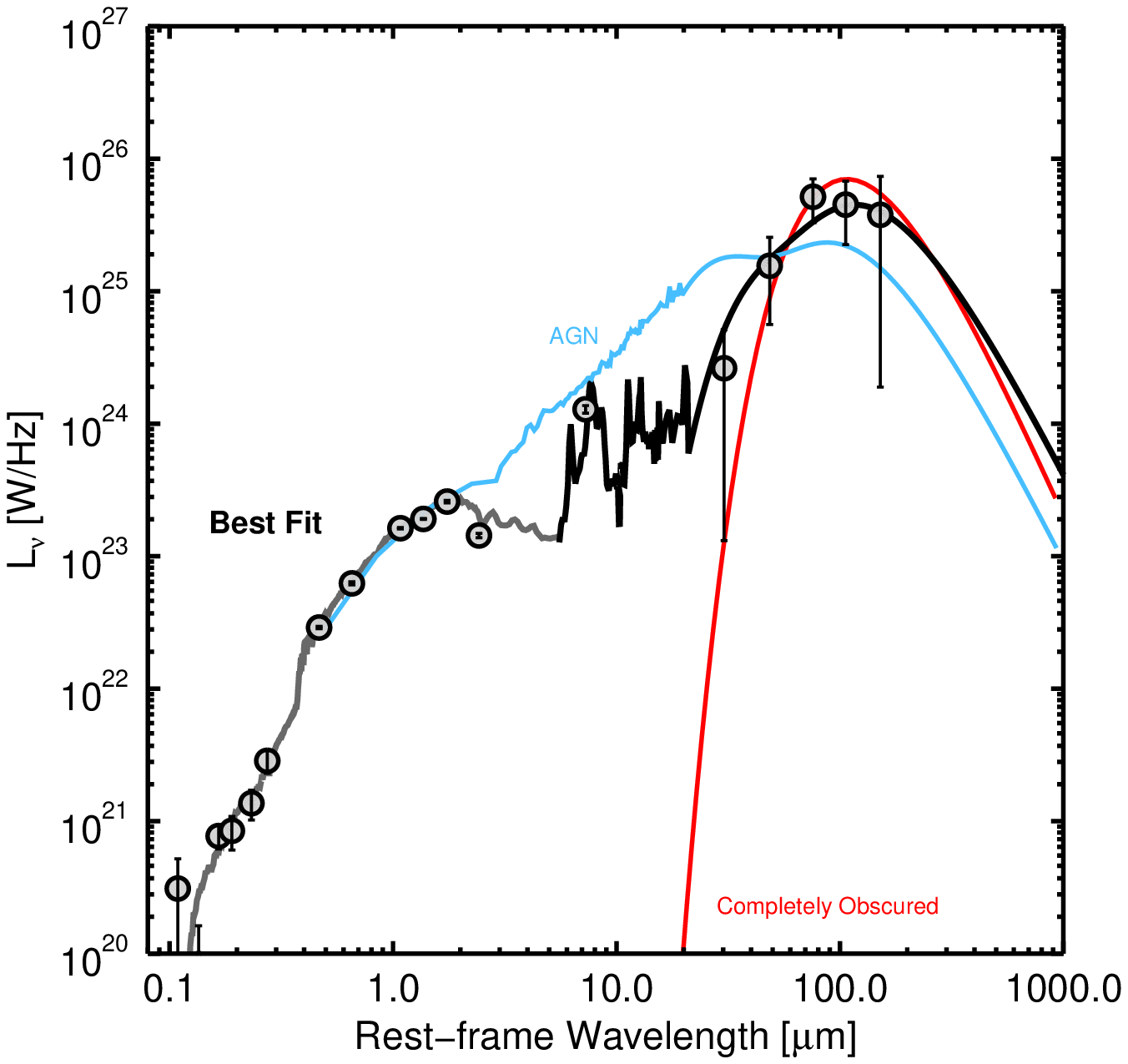}
{\small \bf {\textsf{Figure 3}} $|$ }
{\small \bf {\textsf{UV-FIR spectral energy distribution of GOODS-N-774.}}}
{\small
Rest-frame UV-FIR photometry of GOODS-N-774. Error bars are s.d.
A stellar population synthesis model fit\cite{kriek:09}
to the UV-NIR SED is shown in gray. 
The black line shows the
composite star formation + AGN SED\cite{kirkpatrick:12}
that is the best fit to the mid- and
far-IR data. This best fit has an AGN contribution of 0\,\% and
implies a star formation rate of 90\,\msun/yr.
For reference, the light blue line
shows a composite SED with an AGN contribution of 80\,\%.
The red curve shows
a black body with a size of 1\,kpc and a bolometric luminosity
of $10^{12}$\,L$_{\odot}$.
}
\end{figure}

This mismatch
could imply that the lifetime of the compact star forming 
phase is very short, as has been suggested
previously based on similar number density arguments.\cite{tacconi:08}
It may be that we are witnessing the
aftermath of the contraction of a gravitationally unstable star-forming 
disk\cite{dekel:14} or of a merger of large star forming galaxies \cite{tacconi:08}. 
However, neither tidal features nor extended wings 
are apparent in the surface density distribution.

It is perhaps more likely that the lifetime of the compact star forming phase 
is relatively long and that we are missing many star forming compact
galaxies in current surveys. 
From the compact morphology and high star formation rate we infer a
high gas column density for this object\cite{kennicutt:98}: 
$N_H=2.6\times10^{23}$\,cm$^{-2}$.
This gas column density is nearly an order of magnitude higher 
than in an average UV-selected 
star-forming galaxy at the same cosmic epoch\cite{erb:06} and 
two and a half orders of magnitude higher than
in the disk of a typical galaxy in the local universe\cite{kennicutt:98}.
This high column density of gas in conjunction
with the abundance of metals implies\cite{gilli:14} a very high extinction:
$A_V\gtrsim 100$ for a screen, and $A_V\gtrsim 6$ if the dust and the
stars are mixed. %\cite{bohlin:78} 
The detection of
rest-frame optical flux, and of H$\alpha$ emission, are inconsistent
with such high values. The dust distribution is probably non-uniform
and it may be that, for GOODS-N-774, we are looking
through a relatively unobscured line of sight.

\begin{figure*}[htbp]
\epsfxsize=15cm
\epsffile{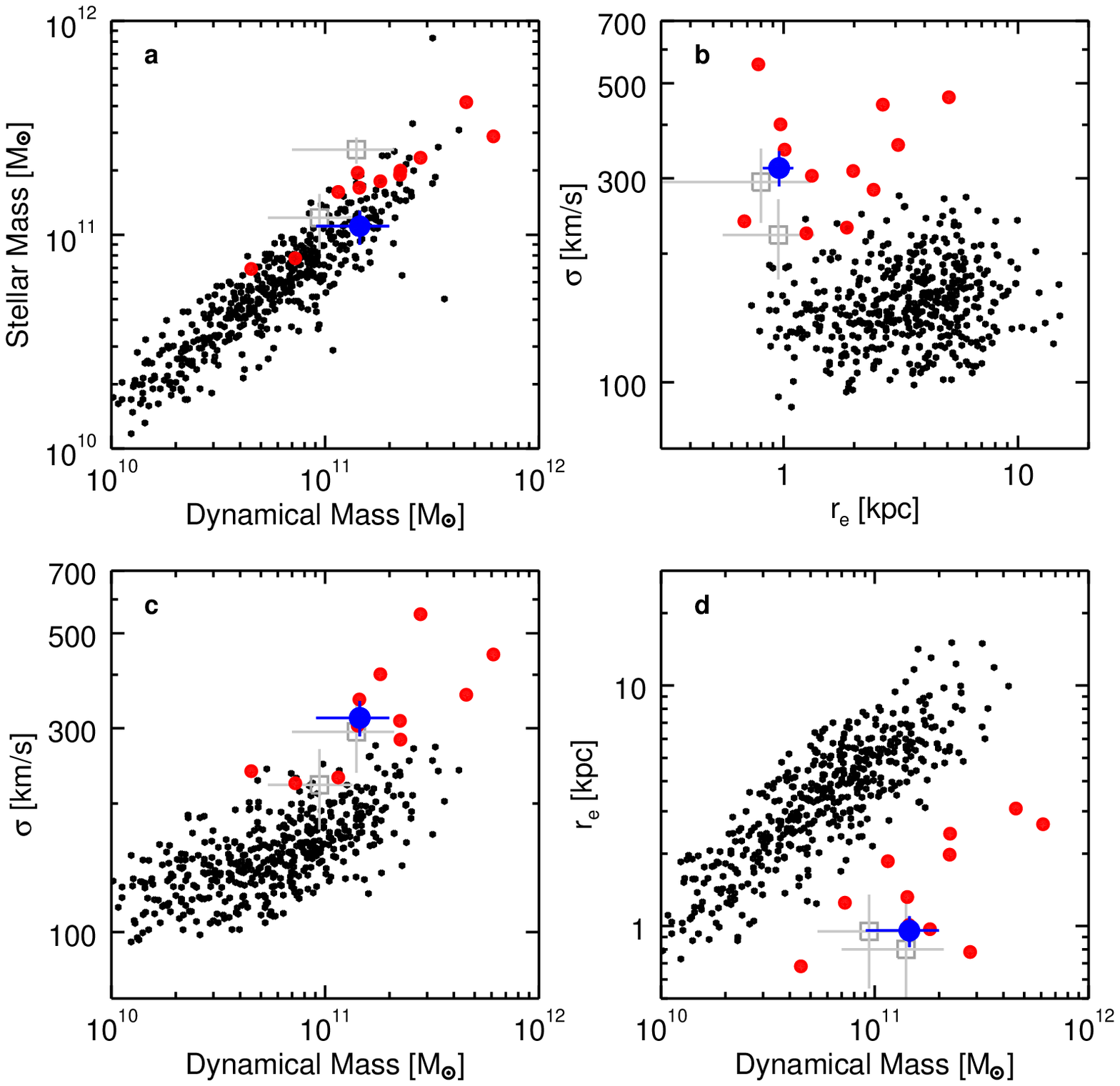}\\
{\small \bf {\textsf{Figure 4}} $|$ }
{\small \bf {\textsf{Properties of GOODS-N-774 compared to quiescent galaxies.}}}
{\small
Panels a-d compare the size, mass, and gas dynamics
of GOODS-N-774 (blue
symbol) to the sizes, masses, and stellar dynamics
of galaxies in SDSS (black) and massive quiescent
galaxies at $z\sim 2$
(red).\cite{bezanson:13,vandokkum:09b,vandesande:13,belli:14}
GOODS-N-774 has properties that are similar to previously studied
massive quiescent galaxies at $z\sim2$ and is substantially offset from nearby
galaxies.
CO dynamics and CO sizes of two compact SMGs 
from ref.\ 13 (HDF 76 and N2850.2) are shown in grey.
Error bars are s.d.
}
\end{figure*}
%\vspace{-0.5cm}

More typical star forming cores could be entirely
obscured,\cite{wang:12,gilli:14} and begin
to resemble
black bodies with a temperature of $\sim 30$\,K (red curve in Fig.\ 3;
calculated using a radius of 1\,kpc and
$L_{\rm bol}=10^{12}$\,L$_{\odot}$).
It may be possible to select such obscured progenitors at long wavelengths, 
near the peak of the redshifted dust emission. It has been demonstrated 
that redshifts, sizes, and velocity widths of IR-luminous galaxies can be 
measured from CO emission. In fact, the closest analogs 
to GOODSN-774 are the
two submm-selected galaxies (SMGs) HDF 76 and N2850.2
(see Fig.\ 4), 
which have high line widths and small sizes in the CO line.\cite{tacconi:08}
It will be 
interesting to determine whether the stellar distribution of these galaxies 
is similar to the gas distribution, or these are dense star forming regions
inside larger galaxies.

Longer wavelength studies of large, unbiased samples
can show whether GOODS-N-774
is an example of a parent population of 
compact star forming galaxies that are heavily
obscured.\cite{tacconi:08}
There may also be multiple paths to a compact, quiescent galaxy: some (such as
HDF 76 and N2850.2\cite{tacconi:08}) may
form most of their stars
in mergers with star formation rates of $\gtrsim 1000$\,\msun/yr, whereas
others (such as GOODS-N-774) may grow relatively slowly in an obscured,
accretion-throttled mode.
Whatever the dominant mode turns out to be,
as the stars in dense cores account for 10\,\% -- 20\,\%
of the total $z\sim 2$ stellar mass density,\cite{vandokkum:14}
star forming cores should account for a significant fraction of
all star formation in the high redshift Universe.
\vspace{0.3cm}\\
Very recently, evidence supporting our conclusions has been 
posted online.\cite{barro:14b}

\begin{figure}
%\hspace{-0.1cm}
\epsfxsize=7.5cm
\epsffile{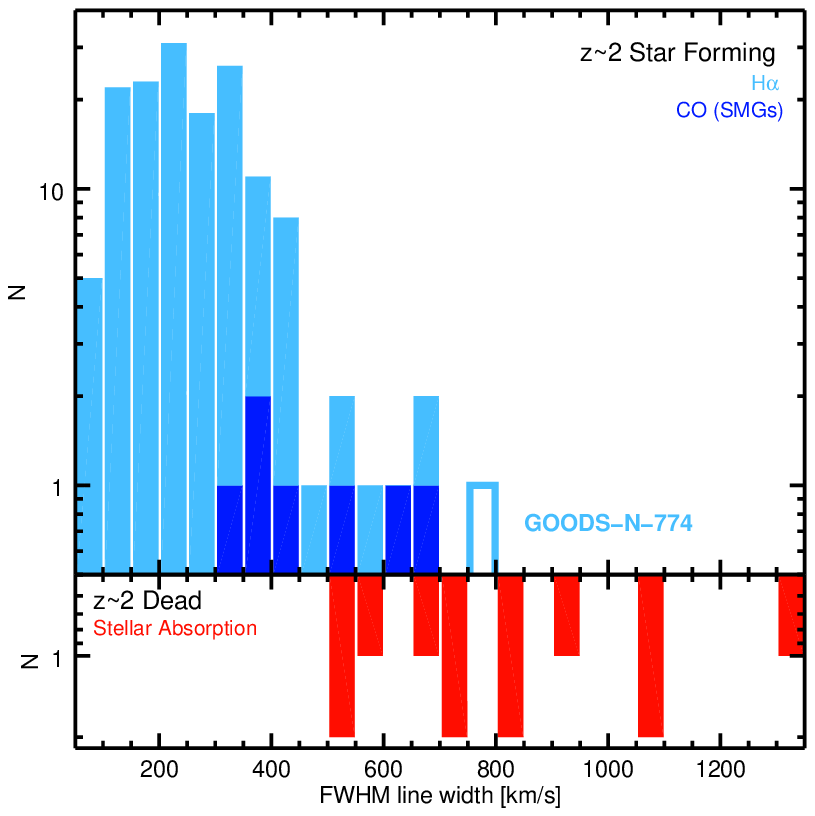}\\
{\small \bf {\textsf{Extended Data Figure 1}} $|$ }
{\small \bf {\textsf{Line widths of $z\sim 2$
star-forming and quiescent galaxies.}}}
{\small
The line width of GOODS-N-774 (open box) is among the highest measured for a
normal star forming galaxy at high redshift in 
 \ha$^{17,18}$(light blue) %\cite{erb:06,forsterschreiber:09}
 or CO emission (SMGs)$^4$ (dark blue).
 The gas velocity dispersion is similar to the median
stellar velocity dispersion of 304\,km/s in 
a sample of quiescent galaxies at
$z=1.5-2.2$ with median mass $1.9\times10^{11}$\msun\,(red).$^{8-11}$
%\cite{bezanson:13,vandokkum:09b,vandesande:13,belli:14}
}
\end{figure}

\vspace{1cm}
\noindent
\noindent\textbf{METHODS}
\vspace{0.05cm}\\
\textbf{Spectral energy distribution.}
The candidate forming core was found using the 3D-HST 
catalogs in the five CANDELS fields. CANDELS is a 902 orbit 
Hubble Space Telescope program that provides
space-based optical and near-infrared imaging across $\sim900$\,arcmin$^2$.
\cite{grogin:11,koekemoer:11} Aperture photometry was performed 
to produce publicly available photometric catalogs and to derive stellar 
masses.$^{12,14}$
24\,$\mu$m fluxes from Spitzer MIPS were determined using the same
methodology as ref.\ 33.
The derived fluxes are consistent 
with the public
catalog of ref.\ 34.
Using the 24\,$\mu$m data as position priors we measure
the 100\,$\mu$m -- 500\,$\mu$m fluxes from the ultra-deep
Herschel imaging in GOODS-North.\cite{elbaz:11}
In sum, the rest-frame UV -- optical data come from HST/ACS, HST/WFC3,
and ground-based optical telescopes;
the rest-frame near-IR data are from Spitzer/IRAC; the mid-IR point
is from Spitzer/MIPS; and the far-IR data are from Herschel/PACS
and SPIRE.
\vspace{0.05cm}\\
\textbf{Keck spectroscopy.}
We observed GOODS-N-774 with the near infrared spectrograph 
(NIRSPEC) on the W.\ M.\ Keck telescope in the $K$ band, on January 11, 2014.
The total integration time was 6000\,s. 
We used the low dispersion mode
with a slit width of $0.7''$, giving a spectral resolution of
$\sigma_{\rm instr}=6.1$\,\AA\ in the rest-frame. 
We fit a Gaussian to the \ha\, $\lambda6563$\AA\, and [N\,{\sc ii}]
$\lambda\lambda6548,6584$\AA\, emission lines simultaneously and
corrected for the instrumental resolution. 
The uncertainty in
the derived properties was determined by refitting the model with
empirical realizations of the noise. 
\vspace{0.05cm}\\
\textbf{HST grism spectroscopy.}
A WFC3/G141 grism
spectrum of the object was obtained as part of the 3D-HST 
survey.\cite{brammer:12}
3D-HST is a near infrared slitless spectroscopic Treasury program.
We examined the grism spectrum after measuring a secure redshift from the
Keck/NIRSPEC spectrum.
The redshifted [O\,{\sc ii}], H$\beta$, and [O\,{\sc iii}]
lines are detected with a significance
of $1.5\sigma - 2.5\sigma$.  
GOODS-N-774 has optical emission line ratios 
[O\,{\sc iii}]/H$\beta=1.2\pm0.9$ and [NII]/\ha\,$=0.4\pm0.1$, 
suggesting a level of gas excitation that is slightly higher 
than the locus of star forming galaxies in the local universe\cite{tremonti:04} 
but at the low end for star forming galaxies at $z\sim2$\cite{steidel:14}
in the diagnostic BPT diagram.
\vspace{0.05cm}\\
\textbf{X-ray constraints.}
GOODS-N-774 is in the Chandra Deep Field North, which has been observed
for a total of $\approx 2$\,Ms
with the Chandra X-ray satellite. The exposure time at the location
of GOODS-N-774 is 1.22\,Ms.
The galaxy is not in the publicly available point-source catalog
of this field.\cite{alexander:03} There are 7 counts in
a $3''$ aperture centered on the object location in
the full band ($0.5-8$\,keV) X-ray image, fully
consistent with the counts in random apertures in regions with
the same exposure time. Using the s.d.\ of the counts in
random apertures we derive a $2\sigma$ upper limit of 6 counts
for the X-ray flux of GOODS-N-774.
Using PIMMS v4.6b we derive a rest-frame $2-10$\,keV flux limit
of $F_X < 2.9\times 10^{-17}$\,ergs\,s$^{-1}$\,cm$^{-2}$, corresponding
to a luminosity $L_X<1.2\times 10^{42}$\,ergs\,s$^{-1}$.
We conclude that there is no evidence for an AGN in GOODS-N-774.
The upper limit is consistent with the star formation rate of
the galaxy.\cite{grimm:03}\\
\vspace{0.05cm}\\
\textbf{Gas column density.}
We derive the gas surface density using the Kennicutt-Schmidt law$^{23}$:
$$\Sigma_{SFR}=(2.5\pm0.7)\times10^{-4}(\frac{\Sigma_{gas}}{1 {\rm M}_\odot {\rm pc}^{-2}})^{1.4\pm0.15}\frac{{\rm M}_\odot}{ {\rm yr}\cdot{\rm kpc^2}}.$$
\vspace{0.05cm}\\
\textbf{Dynamical mass.}
We define dynamical mass as $M_{dyn}=k(n)\sigma^2r_e/G$, with
the constant $k(n)$ depending on the S{\'e}rsic index:
$k(n)=8.87-0.831n+0.0241n^2$.\cite{bertin:02}
GOODS-N-774 has a S{\'e}rsic index $n=2.9$;
the comparison samples of compact quiescent galaxies at $z\sim 2^{1,}$
\cite{trujillo:06,toft:07,vandokkum:08,cimatti:08,newman:10}
and SDSS galaxies with $0.058<z<0.060$ have median $n=3.2$ and $n=4.1$
respectively.

\bibliographystyle{nature-pap}
%\bibliography{journals,refs3}
%\bibliography{../../../all.bib}

\vspace{0.3cm}

\vspace{0.3cm}
\noindent
{\small \bf \textsf{Acknowledgements} 
{\small Support from STScI grant GO-1277 is gratefully acknowledged.}

\vspace{0.3cm}
\noindent
{\small \bf \textsf{Author Contributions}}
{\small {E.J.N.\ obtained the data, led the analysis and
the interpretation, and wrote the manuscript.
P.v.D.\ contributed to the analysis and the interpretation.
M.F.\ contributed to the interpretation. I.M. reduced the WFC3 imaging.
G.B.\ and I.M.\ reduced the grism spectroscopy.
K.W. and R.S. led the photometric analysis. 
All authors commented on the manuscript. 
%{\small \textsf{}}
\vspace{0.5cm}\\
{\small \bf \textsf{Author Information}}
{\small \textsf{Reprints and permissions information
are available at npg.nature.com/reprintsandpermissions.
Correspondence and requests for materials
should be addressed to E.J.N.\ (erica.nelson@yale.edu).}}

\end{document}